\documentclass[pre,aps,twocolumn,showpacs,superscriptaddress]{revtex4}
\usepackage{graphicx,latexsym}
\usepackage{amsmath}
\newcommand{\sect}[2]{{\par\it #1.---}{#2}}
\newcommand{\beq}{\begin{equation}}
\newcommand{\eeq}{\end{equation}}
\newcommand{\bqn}{\begin{eqnarray}}
\newcommand{\eqn}{\end{eqnarray}}
\newcommand{\bqns}{\begin{eqnarray*}}
\newcommand{\eqns}{\end{eqnarray*}}
\newcommand{\bary}{\begin{array}}
\newcommand{\eary}{\end{array}}
\newcommand{\non}{\nonumber}

\usepackage{graphicx}%
\usepackage{amsmath}

\begin{document}
\title{Lewis-Riesenfeld approach to the solutions of Schr\"{o}dinger
equation in the presence of a time-dependent linear potential}

\author{Pi-Gang Luan}
\affiliation{Institute of Optical Sciences, National Central
University, Chung-Li 32054, Taiwan}
\author{Chi-Shung Tang}
\affiliation{Physics Division, National Center for Theoretical
Sciences, P.O. Box 2-131, Hsinchu 30013, Taiwan}
\date{\today}

\begin{abstract}
We reexamine the general solution of a Schr\"{o}dinger equation in
the presence of a time-dependent linear potential in configuration
space based on the Lewis-Riesenfeld framework. For comparison, we
also solve the problem in momentum space and then Fourier transform
the solution to get the general wave function. Appropriately
choosing the weight function in the latter method, we can obtain the
same wave function as the former method. It is found that
non-Hermitian time-dependent linear invariant can be used to obtain
Gaussian-type wave-packet (GTWP) solutions of the time-dependent
system. This operator is a specific linear combination of the
initial momentum and initial position operators. This fact indicates
that the constants of integration such as the initial position and
initial momentum that determine the classical motion play important
roles in the time-dependent quantum system. The eigenfunction of the
linear invariant is interpreted as a wave packet with a ``center of
mass" moving along the classical trajectory, while
the ratio between the coefficients of the initial position and
initial momentum determines the width of the wave packet.

\end{abstract}

\maketitle

\sect{Introduction}
The study of time-dependent systems has been a growing field not
only for its fundamental physical perspective but also for its
applicability, such as quantum transport~\cite{QT1,QT2,QT3}, quantum
optics~\cite{QO1,QO2,QO3}, quantum information~\cite{QI1,QI2}, and
spintronics~\cite{spin1,spin2,spin3}. Recently, there has been
attracted attention of physicists in the analytical solutions of the
one-dimensional Schr\"{o}dinger equation with a time-dependent
linear potential~\cite{Guedes01,MF,Bekkar,JB}. First, Guedes solved
the wave function for a Schr\"{o}dinger equation with a
time-dependent linear potential~\cite{Guedes01}, using the
Lewis-Riesennfeld (LR) method~\cite{Lewis,LR}. Later on, Feng
\cite{MF} found the plane-wave-type and the Airy-packet solutions
using a space-time transformation method. However, Bekkar \textit{et
al.} pointed out that the Airy-packet solution is in fact only a
superposition of the plane-wave-type solution~\cite{Bekkar}.
Moreover, Bauer \cite{JB} showed that the solution proposed by
Ref.~\onlinecite{Guedes01} is only a special case of the so-called
Volkov solution with a zero wave vector $k$. He also showed how to
do the gauge transformation appropriately to the time-dependent
Schr\"{o}dinger equation to get the different expressions of the
Hamiltonian and the corresponding Volkov solution.

Besides the solutions described above,  we shall show that the
linear invariant can be \textit{non-Hermitian}. As such, a GTWP
solution is then derived naturally based on the LR approach. This
solution was ruled out in previous studies because the linear LR
invariant $\hat{I}(t)$ as a \textit{Hermitian} operator had been
assumed~\cite{Guedes01,MF,Bekkar,JB}. Although
in Ref.~\onlinecite{Bekkar} the authors pointed out the
incorrectness of setting $B=0$ in Refs.\onlinecite{Guedes01}, the
Hermiticity assumption still led them to the conclusion of $B=0$,
which they commented as ``a constraint that must be taken to get
physical solutions." This assumption, however, is actually
unnecessary.

The main results of this report are as follows. First, we show that
the GTWP solution is derived using a non-Hermitian linear LR
invariant. Second, we solve the Schr\"{o}dinger equation in momentum
space and then transform back to the coordinate space to obtain a
general wave function solution. Third, we present a physical
interpretation to the LR invariant for the  realization of
constructing an invariant and the selection of an appropriate
space-time transformation to find the general solution.

Schr\"{o}dinger equation for describing the motion of a particle in
the presence of a time-dependent linear potential is of the form
\begin{equation}
i\hbar\frac{\partial\psi}{\partial t}=\hat{H}\psi\label{firsteq}\, ,
\end{equation}
where the Hamiltonian $\hat{H}(t)$ is induced by an external
time-dependent driving force $F(t)$, given by
\begin{equation}
\hat{H}(t)=\frac{\hat{p}^2}{2m}-F(t)\hat{x},
\end{equation}
This time-dependent dynamical problem could be solved in either
configuration or momentum space.

\sect{Configuration space}
To utilize the LR method~\cite{LR} solving the time-dependent
system, one should firstly find an operator $\hat{I}(t)$ such that
\begin{equation}
i\hbar\frac{d \hat{I}}{dt}=i\hbar\frac{\partial \hat{I}}{\partial
t}+[\hat{I},\hat{H}]=0, \label{invariant} \, ,
\end{equation}
and then find its eigenfunction $\varphi_\lambda(x,t)$ satisfying
\begin{equation}
\hat{I}(t)\,\varphi_{\lambda}(x,t)=\lambda\,\varphi_{\lambda}(x,t),\label{eigenphi}
\end{equation}
with $\lambda$ being the corresponding eigenvalue. The general wave
function $\psi(x,t)$ is then written as
\begin{equation}
\psi(x,t)=\int d\lambda\, g(\lambda)\psi_\lambda(x,t),
\label{Lewispsi}
\end{equation}
where $g(\lambda)$ is a weight function for $\lambda$.

The wave function $\psi_\lambda (x,t)$ in Eq. (\ref{Lewispsi}) is
related to $\varphi_\lambda (x,t)$,
\begin{equation} \label{phaseeq}
\psi_\lambda(x,t)=e^{i\alpha_{\lambda}(t)}\,\varphi_{\lambda}(x,t)\,
,
\end{equation}
where $\alpha(t)$ is a function of time only, satisfying
\begin{equation}
\dot{\alpha}_{\lambda}=\varphi^{-1}_{\lambda}
(i{\partial}/{\partial
t}-\hat{H}/\hbar)\varphi_{\lambda}.\label{alpha1}
\end{equation}
We note that the integration in Eq.~(\ref{Lewispsi}) includes, in
general, all possible degeneracies of $\lambda$. It turns out that
the time-dependent invariant operator $\hat{I}(t)$ takes the linear
form~\cite{Guedes01}
\begin{equation}
\hat{I}(t)=A(t)\hat{p}+B(t)\hat{x}+C(t)\, ,
\label{iabc}
\end{equation}
in which $A(t)$, $B(t)$, and $C(t)$ are time-dependent $c$-number
functions to be determined.

The operators describing the equations of motion are given by
\begin{equation}
\frac{d\hat{x}}{dt}=\frac{[\hat{x},\hat{H}]}{i\hbar}=\frac{\hat{p}}{m},\label{dxdt}
\end{equation}
and
\begin{equation}
\frac{d\hat{p}}{dt}=\frac{[\hat{p},\hat{H}]}{i\hbar}=F(t).\label{dpdt}
\end{equation}
By solving the above two equations, the space and momentum operators
can be obtained in terms of initial conditions, given by
\begin{equation}
\hat{x}(t) = \hat{x}(0) + \frac{\hat{p}(0)t + G_1(t)}{m}\label{xt}
\end{equation}
and
\begin{equation}
\hat{p}(t)=\hat{p}(0)+G(t),
\label{pt}
\end{equation}
where $G_1(t)$ and $G(t)$ are defined, respectively, as
\begin{equation}
G_1(t)\equiv\int^t_0G(\tau)d \tau
\label{g1}
\end{equation}
and
\begin{equation}
G(t)\equiv\int^t_0F(\tau)d\tau \, .
\label{g}
\end{equation}

Substituting Eqs.~(\ref{iabc}), (\ref{dxdt}), and (\ref{dpdt}) into
Eq.~(\ref{invariant}), and solving these operator equations, we get
\begin{equation}
A(t)=A_0-\frac{B_0}{m}t,\;\;\;\;\;
B(t)=B_0,\label{a0b0}
\end{equation}
\bqn
C(t)&=&C_0-A_0\int^t_0F(\tau)\,d\tau+\frac{B_0}{m}\int^t_0F(\tau)\tau\,d\tau\non\\
&=&C_0-A(t)\,G(t)-\frac{B_0}{m}G_1(t),
\label{c0}
\eqn
where $A_0$, $B_0$, and $C_0$ are arbitrary complex constants.
Furthermore, substituting Eqs.~(\ref{xt})--(\ref{c0}) into (\ref{iabc}), we find
\begin{equation} \hat{I}(t)=A_0\,\hat{p}(0)+B_0\,\hat{x}(0)+C_0=\hat{I}(0).\end{equation}
In other words, the invariant $\hat{I}$ is precisely the linear
combination of the initial momentum $\hat{p}(0)$ and the initial
position $\hat{x}(0)$ with an arbitrary constant $C_0$.

For the convenience of later discussion, we define $x_c(t)$ and
$p_c(t)$ as the expectation value of $\hat{x}(t)$ and $\hat{p}(t)$
with respect to the wave function $\psi_\lambda(x,t)$, i.e.,
\begin{equation} x_c(t)\equiv\langle
\hat{x}(t)\rangle_\lambda=x_0+\frac{p_0t+G_1(t)}{m}\,
,
\end{equation}
\begin{equation}
p_c(t)\equiv\langle\hat{p}(t)\rangle_\lambda=p_0+G(t)\,
,
\end{equation}
where $x_0=x_c(0)$ and $p_0=p_c(0)$ are,
respectively, the initial position and initial momentum of the
corresponding classical problem.

To find a solution of Eq.(\ref{firsteq}), we have to solve Eq.(\ref{eigenphi}) first.
Note that in Eq.(\ref{phaseeq}) the phase factor $e^{i\alpha(t)}$
is a function of time only, thus $\psi_\lambda(x,t)$ is also
an eigenfunction of $\hat{I}$ with the same eigenvalue $\lambda$. It
turns out that
\bqn \lambda &=& A(t)p_c(t) + B(t)x_c(t) + C(t)\non\\
&=& A_0\,p_0 + B_0\,x_0+C_0.\eqn

By solving Eq.(\ref{eigenphi}), after some algebra, we find
\begin{equation}
\varphi_{\lambda}(x,t)=\exp\left\{\frac{i}{\hbar} \left[\frac{2\left(\lambda-C(t)\right)x
-B_0 x^2}{2A(t)} \right]  \right\}.\label{phisol}
\end{equation}
Substituting Eq.~(\ref{phisol}) into Eq.~(\ref{alpha1}), we obtain
\begin{equation}
\alpha_{\lambda}(t)=\alpha_\lambda(0)-\int^t_0
\left[\frac{(\lambda-C(\tau))^2 + i\hbar B_0 A(\tau)}{2m\hbar
A^2(\tau)}\right] d\tau.
\end{equation}
Here we see that in general $\alpha(t)$ is a complex function.

Using the following identities:
\bqn \frac{\lambda-C(t)}{A(t)}
&=&p_c(t)+\frac{B_0}{A(t)}x_c(t),\non\\
\frac{d}{dt}(B_0/A(t))&=&B^2_0/(mA^2(t)),\non\\
\frac{d}{dt}x^2_c(t)&=&\frac{2}{m}p_c(t)x_c(t), \eqn we finally
obtain the general expression of the wave packet solution
\begin{eqnarray}
\psi_\lambda(x,t)&=&\frac{e^{i\alpha_\lambda(0)}}{\sqrt{A(t)/A_0}}
\exp\left[-\frac{i}{\hbar}\int^t_0\frac{p^2_c(\tau)}{2m}d\tau\right]\non\\
&\times&\exp\left[-\frac{iB_0(x-x_c(t))^2}{2\hbar
A(t)}+\frac{i}{\hbar}p_c(t)x\right].
\label{final}
\end{eqnarray}

It is important to note that since the probability density is of the
form
\begin{equation}
|\psi_\lambda(x,t)|^2= \frac{ e^{ -2\,{\rm Im}(\alpha_\lambda(0)) +
{\rm Im}\left({\cal F}_0\right)
 \frac{\left(x-x_c(t)\right)^2}{\hbar\,\left|A(t)/A_0\right|^2} } }
 {\left|A(t)/A_0\right|}\, ,\label{prob}
\end{equation}
where the factor ${\cal F}_0$ is defined by ${\cal F}_0 \equiv B_0
/A_0$.
It is crucial to note that ${\cal F}_0$ must satisfy
\begin{equation} {\rm
Im}\left({\cal F}_0\right)\leq 0\, .
\end{equation}
to ensure Eq.~(\ref{final}) really providing a physically acceptable
solution. This is a key result of this report.

In case that ${\rm Im}\left({\cal F}_0\right)<0$, the $\psi_\lambda$
obtained in Eq.~(\ref{final}) describes exactly a GTWP with position
uncertainty
\begin{equation}
\Delta x=\sqrt{\frac{\hbar}{2}} \left(\frac{|A(t)/A_0|}{\sqrt{-{\rm
Im}\left( {\cal F}_0 \right)}}\right),\label{dx}
\end{equation}
and momentum uncertainty
\begin{equation}
\Delta p=\sqrt{\frac{\hbar}{2}} \left(\frac{|{\cal
F}_0|}{\sqrt{-{\rm Im}\left({\cal F}_0\right)}}\right),
\label{dp}
\end{equation}
which lead to the uncertainty relation
\begin{equation}
\Delta x\Delta p=\frac{\hbar}{2}\left(\frac{\left|{\cal
F}_0\left(1-{\cal F}_0\frac{t}{m} \right)\right|}{-{\rm
Im}\left({\cal F}_0\right)}\right)\geq\frac{\hbar}{2}.
\end{equation}
The equality holds at $t={\rm Re}\,(m/{\cal F}_0)$. That is, at that
time the position uncertainty of the particle goes to the minimum.

We can see that, in general, $B_0$ (or ${\cal F}_0$) is not zero
although it is taken to be zero in
Ref.~\onlinecite{Guedes01,MF,Bekkar,JB}. In case that ${\rm
Im}({\cal F}_0)=0$, the result obtained in Eq.~(\ref{prob}) enforces
${\cal F}_0=0$. This means that the solution becomes plane-wave-like
rather than GTWP, otherwise $|\psi_\lambda|^2$ becomes divergent at
the moment $t = m / {\cal F}_0$.

\sect{Momentum space}
We now turn to solve the problem in the momentum space by denoting
the wave function as $\phi(p,t)$ and using the substitution
$x\rightarrow i\hbar\,{\partial}/{\partial p}$, we thus have
\begin{equation}
i\hbar\left(\frac{\partial}{\partial t}+F\frac{\partial}{\partial p}\right)\phi
=\frac{p^2}{2m}\phi.\label{linear}
\end{equation}

By changing the momentum and time variables from $(p,t)$ to
$(p',t')$:
\begin{equation}
p'\equiv p-G(t),\;\;\;\;\;t'\equiv t,\label{pptt1}
\end{equation}
the corresponding differential operators can be transformed to the
form
\begin{equation}
\frac{\partial}{\partial p}=\frac{\partial}{\partial p'},\;\;\;\;\;\;\;
\frac{\partial}{\partial t}=\frac{\partial}{\partial t'}
-F(t)\frac{\partial}{\partial p'}.
\label{pptt}
\end{equation}
It turns out that Eq.~(\ref{linear}) becomes
\begin{equation}
i\hbar\frac{\partial\phi}{\partial t'}
=\frac{\left[p'+G(t')\right]^2}{2m}\phi,
\end{equation}
which yields
\begin{equation}
\phi(p',t')=\phi_0(p')
\exp\left\{-\frac{i}{\hbar}\int^{t'}_0\frac{\left[p'+G(\tau)\right]^2}{2m}\,d\tau\right\},
\end{equation}
or equivalently
\bqn \phi(p,t)&=&\phi_0\left[p-G(t)\right]\non\\
&\times&\exp\left\{-\frac{i}{\hbar}\int^t_0\frac{\left[p-G(t)+G(\tau)\right]^2}{2m}d\tau\right\}.
\label{gen}\eqn
Here $\phi_0$ is an arbitrary single-variable function.

The general solution of the wave function $\psi(x,t)$ can be obtained
by the Fourier transform:
\bqn \psi(x,t)&=&\frac{1}{\sqrt{2\pi\hbar}}\int^{\infty}_{-\infty} \phi(p,t)\,e^{ipx/\hbar}dp,\non\\
&=&\frac{1}{\sqrt{2\pi\hbar}}\int^{\infty}_{-\infty}\phi_0(p')\exp\left\{\frac{i}{\hbar}
\left[p'+G(t)\right]x\right.\non\\
&&\left.-\frac{i}{\hbar}\int^t_0\frac{[p'+G(\tau)]^2}{2m}d\tau\right\}dp'.\label{pg}
\eqn

\sect{Comparison}
We now show that the GTWP solution Eq.~(\ref{final}) obtained using
LR method can also be obtained from the general solution (\ref{pg}).
We consider the time-dependent wave function $\phi_0(p')$ in
momentum space, given by
\begin{equation}
\phi_0(p')=\left(\frac{2\sigma^2}{\pi\hbar^2}\right)^{\frac{1}{4}}
\exp\left[-\frac{\sigma^2(p'-p_0)^2}
{\hbar^2}-i\frac{(p'-p_0)x_0}{\hbar}\right].
\label{gauss}
\end{equation}
After some algebra, we obtain the space-time wave function, given
explicitly
\bqn
\phi(p,t)&=&\left(\frac{2\sigma^2}{\pi\hbar^2}\right)^{\frac{1}{4}}
\exp\left[-\frac{i}{\hbar}\int^t_0\frac{p^2_c(\tau)}{2m} d\tau\right]\non\\
&\times&\exp\left[-\frac{\sigma^2(1+it/T)}{\hbar^2}\left(p-p_c(t)\right)^2\right]\non\\
&\times&\exp\left[-\frac{i}{\hbar}\left(p-p_c(t)\right)x_c(t)\right].\label{plong}
\eqn
Here the paramter
\begin{equation}
T\equiv\frac{2m\sigma^2}{\hbar}
\end{equation}
indicates a measure of the \textit{spreading time} of the GTWP.

Substituting Eq.~(\ref{gauss}) or Eq.~(\ref{plong}) into Eq.~(\ref{pg})
and accomplishing the integration, we get \bqn
\psi(x,t)&=&\frac{1}{(2\pi)^{\frac{1}{4}}\sqrt{\sigma(1+it/T)}}
\exp\left[-\frac{i}{\hbar}\int^t_0\frac{p^2_c(\tau)}{2m} d\tau\right]\non\\
&\times&\exp\left[-\frac{\left(x-x_c(t)\right)^2
}{4\sigma^2(1+it/T)}+\frac{i}{\hbar}p_c(t)x\right].\eqn By suitably
choosing the initial condition parameters as
\begin{equation} e^{i\alpha_\lambda(0)}=\frac{1}{(2\pi\sigma^2)^{\frac{1}{4}}},\;\;\;\;\;\;
\frac{B_0}{A_0}= -\frac{im}{T}\, ,
\end{equation}
we can see that this wave function is exactly of the form given by
Eq.~(\ref{final}).

\sect{Summary}
In this report we have studied the Schr\"{o}dinger equation with a
time-dependent linear potential. We reexamine the linear invariant
proposed by Guedes~\cite{Guedes01}. We have shown that if we assume
this opterator to be a non-Hermitian one, then a GTWP solution can
be obtained. This GTWP has a ``centor of mass" moving along the
trajectory of the corresponding classical particle. The trajectory
is determined by the classical initial position $x_0$ and initial
momentum $p_0$. In the corresponding quantum problem, $\hat{x}(0)$
does not commute with $\hat{p}(0)$, thus the particle described by a
wave function propotional to the eigenfunction of the linear
invariant operator $\hat{I}=A_0\hat{p}(0)+B_0\hat{x}(0)+C_0$
acquires position and momentum uncertainties, as given by
Eqs.(\ref{dx}) and (\ref{dp}), respectively. The size of the
uncertainties are determined by the ratio ${\cal F}\equiv B_0/A_0$.

On the other hand, we have investigated the time-dependent system in
momentum space. After performing transformation of variables, the
problem becomes exactly solvable. Moreover, we have presented a
specific example for comparison between these two approaches. Our
analysis has shown that the key of solving the time-dependent
Schr\"{o}dinger equation is to find a way to transform the problem
to a standard form. For a linear time-dependent case, the standard
form is simply a free-particle problem. It is interesting to note
that if we treat the driving force as a time-dependent gravity, then
an observer in the ``free-fall frame" will not be able to feel the
gravity. As a result, the frame effectively becomes an inertial
frame. This provides a physical picture for the transformation
Eq.~(\ref{pptt1}) we have performed.

\sect{Acknowledgment}
We would like to thank D.H. Lin and Y.M. Kao for discussion of the
results. This work was partly supported by the National Science
Council, by the National Center for Theoretical Sciences, and by the
National Central University in Taiwan.

\end{document}